\documentclass[aps,prd,twocolumn,superscriptaddress,nofootinbib,preprintnumbers]{revtex4-1}
\usepackage{amsmath}
\usepackage{graphicx}
\usepackage{subfigure}
\usepackage{amssymb}
\usepackage{xcolor}
\usepackage{multirow}
\usepackage{cancel}
\usepackage{color}
\usepackage{ulem}
\usepackage{cases}
\usepackage{listings}
\usepackage{xcolor}
\usepackage{float}
\usepackage{slashed}

\usepackage{tikz}
\usetikzlibrary{arrows,shapes}
\usetikzlibrary{trees}
\usetikzlibrary{matrix,arrows} 				
\usetikzlibrary{positioning}				
\usetikzlibrary{calc,through}				
\usetikzlibrary{decorations.pathreplacing}  
\usepackage{pgffor}							

\usetikzlibrary{decorations.pathmorphing}	
\usetikzlibrary{decorations.markings}
\tikzset{
    vector/.style={decorate, decoration={snake}, draw},
	provector/.style={decorate, decoration={snake,amplitude=2.5pt}, draw},
	antivector/.style={decorate, decoration={snake,amplitude=-2.5pt}, draw},
    fermion/.style={draw=black, postaction={decorate},
        decoration={markings,mark=at position .55 with {\arrow[draw=black]{>}}}},
    fermionbar/.style={draw=black, postaction={decorate},
        decoration={markings,mark=at position .55 with {\arrow[draw=black]{<}}}},
    fermionnoarrow/.style={draw=black},
    gluon/.style={decorate, draw=black,
        decoration={coil,amplitude=4pt, segment length=5pt}},
    scalar/.style={dashed,draw=black, postaction={decorate},
        decoration={markings,mark=at position .55 with {\arrow[draw=black]{>}}}},
    scalarbar/.style={dashed,draw=black, postaction={decorate},
        decoration={markings,mark=at position .55 with {\arrow[draw=black]{<}}}},
    scalarnoarrow/.style={dashed,draw=black},
    electron/.style={draw=black, postaction={decorate},
        decoration={markings,mark=at position .55 with {\arrow[draw=black]{>}}}},
	bigvector/.style={decorate, decoration={snake,amplitude=4pt}, draw},
}

\tikzstyle{block} = [draw, rectangle, 
    minimum height=3em, minimum width=6em]

\usepackage[colorlinks,citecolor=blue]{hyperref}
\usepackage{amsmath}
\usepackage{wrapfig}

\newcommand{\aln}[1]{\begin{align}#1\end{align}}
\newcommand{\be}{\begin{equation}}
\newcommand{\ee}{\end{equation}}
\newcommand{\beq}{\begin{equation}}
\newcommand{\eeq}{\end{equation}}
\newcommand{\bea}{\begin{eqnarray}}
\newcommand{\eea}{\end{eqnarray}}
\newcommand{\beqa}{\begin{eqnarray}}
\newcommand{\eeqa}{\end{eqnarray}}
\newcommand{\besp}{\begin{equation}\begin{split}}
\newcommand{\eesp}{\end{split}\end{equation}}

\def\blue#1{{\textcolor{blue}{#1}}} 

\newcommand{\Dfbd}{\mathord{\buildrel{\lower3pt\hbox{$\scriptscriptstyle\leftrightarrow$}}\over {D}_{\mu}}}

\def\0{\textbf{0}}
\def\1{\textbf{1}}
\def\2{\textbf{2}}
\def\3{\textbf{3}}
\def\4{\textbf{4}}
\def\5{\textbf{5}}
\def\6{\textbf{6}}
\def\7{\textbf{7}}
\def\8{\textbf{8}}
\def\9{\textbf{9}}

\begin{document}
\preprint{YITP-22-115}
\title{Entropy and its conservation in expanding  Universe}

\author{Sinya Aoki}
\email{saoki@yukawa.kyoto-u.ac.jp}
\affiliation{Center for Gravitational Physics and Quantum Information, Yukawa Institute for Theoretical Physics, Kyoto University, Kitashirakawa Oiwakecho, Sakyo-ku, Kyoto 606-8502, Japan}

\author{Kiyoharu Kawana}
\email{kkiyoharu@kias.re.kr}
\affiliation{School of Physics, KIAS, Seoul 02455, Korea}

\begin{abstract}
We investigate properties of the conserved charge in general relativity, recently proposed by one of the present authors with his collaborators,
in the inflation era, the matter dominated era and the radiation dominated era of the expanding Universe. 
We show that the conserved charge in the inflation era becomes the Bekenstein-Hawking entropy for de Sitter space, and it becomes the matter entropy and the radiation entropy in the matter and radiation dominated eras, respectively,
while the charge itself is always conserved. 
These properties are qualitatively confirmed by a numerical analysis of a model with a scalar field and radiations. 
Results in this paper provide more evidences on the interpretation that the conserved charge in general relativity corresponds to entropy.
\end{abstract}

\maketitle

\section{Introduction}

Entropy is one of the most fundamental quantities in nature and plays an important role to connect  microscopic and macroscopic physics. 
For example, in the micro-canonical statistical mechanics, it is given by $S(E,V,N)=\ln\Omega(E,V,N)$ where $\Omega(E,V,N)$ is a number of micro states
in a system having definite $E, V, N$, suggesting that entropy is defined from some microscopic degree of freedom (dof). 
In thermodynamic limit $V\rightarrow \infty$, this entropy correctly reproduces various properties of macroscopic thermodynamic entropy. 
%

%
Another famous example is of course a black hole~\cite{Bekenstein:1973ur,Hawking:1975vcx,Bekenstein:1974ax}, whose entropy is given by 
\be
S_H^{}={A_H \over 4G_N },
\label{BH entropy}
\ee
where $A_H^{}$ is an area of the black hole horizon and $G_N^{}$ is the Newton constant. 
This expression suggests that the microscopic dof of a black hole may be localized at its surface just like (mem)brane objects~\cite{Strominger:1996sh,Maldacena:1999bp,Thorne:1986iy}. 
Although its microscopic origin is still not fully understood yet, 
this formula seems a key to quantum gravity.   
 
In recent works~\cite{Aoki:2020prb,Aoki:2020nzm,Aoki:2022gez,Aoki:2022ugd}, one of the present authors  and his collaborators proposed a new definition of a conserved charge $Q[\xi]$ for a generic metric field $g_{\mu\nu}^{}$, even when a system has no Killing vectors.  (See Section~\ref{sec:review} for details.) 
In particular, it was found for several cases that the conserved charge constructed by a particular vector field $\xi^\mu=-\beta \delta^{\mu}_0$ corresponds to entropy of a system with $\beta$ proportional to an inverse temperature. 
In fact, they explicitly showed that the new conserved charge correctly reproduces the Bekenstein-Hawking entropy Eq.~(\ref{BH entropy}) for several black hole systems such as the Schwarzschild black hole and the BTZ blackhole~\cite{Aoki:2020nzm}.
Interestingly, their definition of ``entropy" contains the volume integral of the energy-momentum tensor and the entropy density is localized at the singularity not at the surface.  
This may also open a new insight for the black hole information paradox~\cite{Hawking:1976ra,Hawking:1975vcx,Raju:2020smc}.    
 
\

One of trivial but important aspects of $Q[\zeta]$ is that it is always conserved no matter what happens during the dynamical evolution of a system.  
In particular, it is quite common in cosmology that dominated matter content of the Universe is changing from inflation era to radiation (matter) dominated era. 
If the interpretation of $Q[\zeta]$ as entropy is actually true, it should represent some ``entropy" for each different eras.  
A purpose of this paper is to clarify the physical meaning of $Q[\zeta]$ in such dynamical transition processes in the expanding Universe. 
After a brief review on the new proposal for the conserved charge in general relativity in the next section, we first analytically study a few transition processes (inflation $\to$ matter $\to$ radiation)
and find that $Q[\zeta]$ can be actually interpreted as entropy for each epoch in the following way in Sec.~\ref{sec:model}:   
During the inflation era, $Q[\zeta]$ reproduces the Bekenstein-Hawking entropy for de Sitter space~\cite{Gibbons:1977mu} by taking $\beta_0^{}$ (an initial value of $\beta$) equal to the inverse de Sitter temperature. 
On the other hand, $Q[\zeta]$ simply corresponds to the total matter entropy as $Q[\zeta]=E/T_M$ with a constant matter temperature (energy)  $T_M^{}$ in the matter dominated era, while $Q[\zeta]$ agrees with the radiation entropy in the radiation dominated era taking ${d\over d-1}\beta^{-1}(x^0):=T_R^{}(x^0)$ as radiation temperature.   
In particular, it is notable that the radiation energy density satisfies the Stefan-Boltzmann law  $\rho_R^{}\propto \beta^{-d}$ without assuming thermal equilibrium of radiations.       
While $\beta$ is continuous, the temperature is discontinuous at transition points (from inflation to matter and from matter to radiation).

After the analytical studies in Sec.~\ref{eq:numerical}, we numerically solve a dynamics of a scalar field theory, which emulate the above dynamics, to confirm the conservation of $Q[\zeta]$ and above interpretations.   
In Sec.~\ref{sec:summary}, we conclude our paper.

\newpage

\section{A matter conserved charge}\label{sec:review}

In this section, we briefly review a new definition on a matter conserved charge in general relativity(GR)  proposed in Refs.~\cite{Aoki:2020prb,Aoki:2020nzm,Aoki:2022gez}.

We first define a matter energy in a $d$-dimensional curve spacetime as
\beqa
E(x^0) &:=& \int_{\Sigma(x^0)} [d^{d-1} x]_\mu\, T^\mu{}_\nu (x) \xi_E^\nu (x)~, 
\eeqa
where $T^\mu{}_\nu$ is a (matter) energy momentum tensor (EMT), $\Sigma(x^0)$ is a constant $x^0$ (space-like) hyper-surface with a $d-1$ dimensional hyper-surafce element $ [d^{d-1} x]_\mu$, and $\xi_E^\mu(x) = -\delta_0^\mu$ is a generator of the time translation.
Here we take a coordinate system which satisfies $g_{0j} =0$ for $j=1,2,\cdots, d-1$ without loss of generality.
This energy is not guaranteed to be conserved in general,  except special cases such that $\xi_E$ becomes a Killing vector~\cite{Aoki:2020prb,Aoki:2020nzm,Aoki:2022gez}.

It has been pointed out~\cite{Aoki:2020nzm}, however, that  there always exists a conserved charge in GR, given by
\beqa
Q(x^0) &:=& \int_{\Sigma(x^0)} [d^{d-1} x]_\mu\, T^\mu{}_\nu (x) \beta(x) n^\nu (x)~, 
\eeqa
 where the scalar function $\beta$ is chosen to satisfy 
$\nabla_\mu (T^\mu{}_\nu  \beta n^\nu ) =0$, and $n^\mu$ is a time-like vector.
While an appropriate choice of $n^\mu$ is needed to be clarified for a general case in future studies, it is reasonable to take $n^\mu=\xi^\mu_E$ in this paper as
we will see.
The conservation condition $\nabla_\mu (T^\mu{}_\nu  \beta n^\nu ) =0$ implies
\beqa
Q(x^0_a) -Q(x_b^0) =\int_{x^0_a}^{x^0_b} dx^0\, \int [d^{d-2} x]_i T^i{}_\nu(x)\beta(x) n^\nu(x),~~~~
\eeqa
so that $Q(x^0)$ is $x^0$-independent if contributions at spatial boundaries in the right-hand side vanish.
An absence of such boundary contributions is obviously true if $T^\mu{}_\nu =0$ at the boundary. 
In the presence of non-zero EMT \blue{at} spatial boundaries, $[d^{d-2}x]_i T^i{}_\nu =0$ is always satisfied: If  $[d^{d-2}x]_i T^i{}_\nu \not=0$, the spacetime must develop in the normal direction to the boundary according to the Einstein equation.  It is impossible since there is no spacetime beyond the boundary. 
%
 %

Using $\nabla_\mu^{} T^\mu{}_\nu =0$, the condition for $\beta$ becomes 
\beqa
T^0{}_0 \partial_0 \beta + T^j{}_0 \partial_j \beta + T^\mu{}_\nu \Gamma^\nu_{\mu 0} \beta = 0~, 
\label{eq:con_beta}
\eeqa
which is a linear partial differential equation, whose solution can be easily obtained once an initial condition of $\beta$ is given
at some $x^0$~\cite{Aoki:2020nzm}. 
This charge can be regarded as a conserved charge from the Noether's 1st theorem for a global symmetry of the matter action in GR, called a hidden (global) matter symmetry~\cite{Aoki:2022ugd}. 

In Ref.~\cite{Aoki:2020nzm}, 
it has been pointed out that $Q$ can be interpreted as entropy with $\beta(x)$ identified as a (time-dependent local) inverse temperature,
since they satisfy the first law of thermodynamics for some cases including the expanding Universe.
Furthermore, it was shown that $Q$ also reproduces the Bekenstein-Hawking entropy Eq.~(\ref{BH entropy}) 
for Schwarzschild blackhole and BTZ blackhole~\cite{Aoki:2020nzm}.

In the next section, we will investigate the conserved charge $Q$ for the expanding Universe in more detail,
in order to add more evidences for our interpretation that $Q$ is entropy and $\beta(x)$ is an inverse temperature. 

\section{Conservation in expanding Universe}\label{sec:expanding Universe}
\label{sec:model}
\subsection{Setup}
We consider the expanding Universe described by
\beqa
ds^2=-(dx^0)^{2}+a^2(x^0)\delta_{ij}^{} dx^i dx^j~, 
\eeqa
and the EMT for a perfect fluid as  
\beqa
T^0{}_0=-\rho(x^0)~,\quad T^i{}_j=\delta^i_j P(x^0):=\delta^i_j \omega(x^0) \rho(x^0)~,~~~
\eeqa
whose covariant conservation implies
\beqa
\dot \rho &=& -(d-1)(\rho+P) H = -(d-1)(1+\omega)\rho H~, 
\label{eq:EMT_con}
\eeqa 
where $\omega(x^0)$ is an equation of state, and  $H=\dot{a}/a$.  
Throughout this letter, we take $c=\hbar = k_B=1$.
In this setup, the condition \eqref{eq:con_beta} becomes 
\beqa
{\dot{\beta}\over \beta} &=&(d-1) H{P 
\over \rho} ~.
\label{eq:beta_con}
\eeqa
Energy and charge are evaluated as
\beqa
E(x^0) &=& -\int_{\Sigma_{x^0}} d^{d-1}x\, T^0{}_0 = V_{d-1} [\rho a^{d-1}](x^0), \\
Q(x^0) &=&  -\int_{\Sigma_{x^0}} d^{d-1}x\, T^0{}_0 \beta = V_{d-1} [\rho a^{d-1}\beta](x^0),~~~
\eeqa
where $V_{d-1} = \int d^{d-1}x $ is a co-moving volume.
While the energy is not conserved ($\dot E\not=0$), it is easy to see from eq.~\eqref{eq:EMT_con} and eq.~\eqref{eq:beta_con}
that the charge is indeed conserved ($\dot Q = 0$ ).
Furthermore, using $Q = E\beta$ and eq.~\eqref{eq:beta_con}, we have
\beqa
{d Q\over dx^0} &=& {d E \over dx^0} \beta + E {d\beta\over d x^0} = \left( {d E \over dx^0}  + P  {d V \over dx^0} \right)\beta,
\eeqa
where $V:= V_{d-1} a^{d-1}$ is a space volume. The 1st law of thermodynamics, $T dS = dE + P dV$, suggests that 
$Q$ and $\beta$ may be regarded as entropy and a (time-dependent) inverse temperature, respectively~\cite{Aoki:2020nzm}.

\subsection{Solution for a constant $\omega$}
When $\omega=$ constant, Einstein equation reads
\beqa
\dot H &=& -{(d-1)(1+\omega)\over 2} H^2~,
\eeqa
whose solution is given by
\beqa
H(x^0) &=& {H_0\over 1+ C_0 H_0 x^0}~, \quad C_0:= {(d-1)(1+\omega)\over 2}~.
\eeqa
We then obtain 
\beqa
a(x^0) &=& a_0 (1+ C_0 H_0 x^0)^{1/C_0}~, \\
\rho(x^0) &=& {\rho_0 \over (1+ C_0 H_0 x^0)^2}=\rho_0^{}\left(\frac{a_0^{}}{a(x^0)}\right)^{(d-1)(1+\omega)}~, \\
\beta(x^0) &=& \beta_0  (1+ C_0 H_0 x^0)^{2\omega\over 1+\omega}=\beta_0^{}\left(\frac{a(x^0)}{a_0^{}}\right)^{(d-1)\omega}~,
\eeqa
where $H_0$, $a_0$, $\rho_0$ and $\beta_0$ are initial values. 
Since $Q$ is a conserved charge, it is determined by its initial value as 
\beqa
Q &=&V_{d-1}^{} \rho_0 a_0^{d-1} \beta_0^{}~. 
\eeqa
In the $\omega+1\to 0~(C_0^{}\rightarrow 0)$ limit, we have $H(x^0)=H_0$, $\rho(x^0)=\rho_0$, 
\beqa
a(x^0) =a_0 e^{H_0 x^0}, \quad \beta(x^0) =\beta_0 e^{-(d-1) H_0 x^0}.
\eeqa

\subsection{Model of expanding Universe}
We consider a simplified model of the expanding Universe, whose time evaluation is given as follows: 
(1) Inflation era at $0 \le x^0 < t_M$ where $P=-\rho$ ($\omega=-1$),
(2) Matter dominated era at $t_M \le x^0 < t_R$ where $P=0$ ($\omega=0$), and (3) Radiation dominated era at $t_R \le t$ where $P=\rho/(d-1)$ ($\omega=1/(d-1)$). 
This history captures the essence of inflaton dynamics after the inflation.    
We determine the time dependences of $a, H,\rho, \beta, E$ and $Q$ at each epochs, and fix initial constants $H_0^{}$, $a_0^{}$, $\rho_0^{}$ and $\beta_0^{}$ by requiring that $Q$ represents the de-Sitter entropy during the inflation.

\subsubsection{Inflation era}
Solutions with $\omega=-1$ read
\beqa
H(x^0) &=&H_I, \ \rho(x^0)=\rho_I ={(d-1)(d-2)\over 16\pi G_N} H_I^2,\\
a(x^0) &=&a_0 e^{H_I x^0}, \ \beta(x^0) =\beta_0 e^{-(d-1)H_I x^0},\\
E(x^0) &=& V_{d-1} \rho_I a_0^{d-1} e^{(d-1)H_I x^0}, Q =  a_0^{d-1}V_{d-1}^{} \rho_I \beta_0.~~~~~~
\eeqa
While the energy is exponentially increasing, $Q$ is constant.

As for the volume factor $a_0^{d-1}V_{d-1}^{}$, we can relate it to the total number of the Hubble patches $N_H^{}$ in the following way. 
First, the maximal length light would travel from $x^0=0$ to $x^0=\infty$~\footnote{Here, the choice of initial time $x^0=t_0^{}$ is arbitrary, and $Q$ does not depend on it.   
 }
is determined by  $ds^2=0\leftrightarrow dr/dx^0 =1/a(x^0)= e^{-H_I x^0}/a_0$, which leads to
 \beqa
 r_{\rm max} &=& \int_0^{\infty} {dr\over dx^0} dx^0 ={1 \over a_0 H_I}.
 \eeqa
This means that the Hubble volume is given by
\beqa
V_{H}^{} &:=& a_0^{d-1}\Omega_{d-2} \int_0^{r_{\rm max}}r^{d-2} dr = 
{H_{I}^{-(d-1)}\over d-1} \Omega_{d-2}~,
\eeqa
where $\Omega_{d-2}$ is a solid angle of $d-2$ dimensional polar coordinates. 
Then, the total number of the Hubble patches $N_H^{}$ is 
\aln{N_H^{}=\frac{a_0^{d-1}V_{d-1}^{}}{V_H^{}}~.
}
%
By using this relation, we can rewrite $Q$ as 
 \beqa
 Q &=&N_H^{}\times {A_H\over 4 G_N} \times {(d-2)\over 2} T_H^{}\beta_0~,\\ 
  A_H^{} :&=& R_H^{d-2} \Omega_{d-2}~,
 \eeqa
 where $A_H^{}$ is an area of the Hubble horizon, which is equivalent to the cosmological horizon for de Sitter spacetime with a radius $R_H =1/H_I$, since the metric with $a(x^0)=a_0 e^{H_I x^0}$ describes (a part of) de Sitter spacetime, and thus $T_H^{}=H_I/(2\pi)$ is a temperature of Hubble or de Sitter horizon.  
 Under this equivalence,  the constant $\rho_I$ for $w=-1$ is identified with the  cosmological constant as $\Lambda_{dS} =(d-1)(d-2)/(2R_H^2) = 8\pi G_N \rho_I$,
 which implies $R_H=1/H_I$. 
 Thus, taking an initial condition of $\beta_0$ as
\beqa
 \beta_0^{} ={1\over \tilde T_H^{}}, \quad \tilde T_H^{} := {(d-2)\over 2} T_H^{}~,
 \label{eq:temp_dS}
 \eeqa
and focusing on one Hubble patch i.e. $N_H^{}=1$, $Q$ reproduces the Bekenstein-Hawking entropy formula~\cite{Bekenstein:1973ur,Bekenstein:1974ax,Hawking:1975vcx} for de Sitter spacetime~\cite{Gibbons:1977mu} as $Q = {A_H^{}\over 4 G_N^{}}$.
This fact justifies our interpretation that the conserved charge $Q$ is entropy and $\beta(x^0)$ is proportional to an inverse temperature, 
so that the temperature during the inflation era is given by $T_H^{}(x^0) = {2\over (d-2)\beta(x^0)}$. 
In other words,  we can derive the Bekenstein-Hawking entropy formula for a (static) de Sitter spacetime~\cite{Gibbons:1977mu} from the entropy $Q(x^0)$ in the inflation era with an appropriate choice of the initial inverse temperature $\beta_0^{}$. 
Note however that the entropy $Q$ is uniformly distributed inside the Hubble horizon but is not concentrated on the horizon.  
An overall factor $(d-2)/2$ for $T_H^{}$ in $\beta_0^{}$ becomes unity at $d=4$ and this factor also appears in the Schwarzschild blackhole~\cite{Aoki:2020nzm}.

\subsubsection{Matter dominated era}
 At $x^0 = t_M$, the inflation era ends, 
  and the matter dominated era starts typically due to the oscillations of inflaton around the origin.  

 Solutions with $\omega=0$ lead to time-dependent $H$ and $\rho$ during the matter dominated era as
 \beqa
 H(x^0) &=& H_I^{}\left(\frac{a(t_M^{})}{a(x^0)}\right)^{\frac{d-1}{2}},~ 
 \rho(x^0) = \rho_I^{}\left(\frac{a(t_M^{})}{a(x^0)}\right)^{d-1},~~~~~
 \eeqa
 while  $\beta(x^0) =\beta(t_M):=T_M^{-1}$, $Q$ and 
 \beqa
 E(x^0) = V_{d-1}\rho_I a_0^{d-1} e^{(d-1) H_I t_M}=V_{d-1}^{}\rho_I^{}a(t_M^{})^{d-1}~~~
 \eeqa
 are all constant during the matter dominated era.
Using the law of equipartition, the energy can be also expressed as 
 \beqa
E(t_M^{}) &=& {g_M N_M\over 2} \varepsilon_M^{}~,  
\eeqa
where $N_M$ is a number of matter particles, $g_M$ is an effective degrees of freedom for a particle, and 
$\varepsilon_M$ is the effective temperature (energy) of matter particle.   
Thus a number of particle $N_M$ does not change  during the matter dominated era. 
While $\varepsilon_M^{}$ is different from $T_M^{}$ in general,  
$t_M^{}$ and $T_M^{}$ are related as
\beqa
 t_M &=& {1\over (d-1) H_I} \log {T_M\over \tilde T_H}~,
\eeqa 
where a positivity of $t_M >0$ implies  $T_M > \tilde T_H$, which is easily satisfied, since the temperature increases during the inflation. 
Note also that while $\beta(x^0)$ is continuous at $x^0= t_M$, the temperature $T_H(t_M)$ at the end of the inflation era  and the matter temperature $T_M$ are different 
as $T_M ={(d-2)\over 2} T_H(t_M)$, due to the new relation $T_M=1/\beta(t_M^{}) $, which is required for $Q(t_M)$ to be the matter entropy during the matter dominated era.  

 \subsubsection{Radiation dominated era}
 After $x^0=t_R^{}$, the radiation era starts.  
 Solutions with $\omega=1/(d-1)$ are 
 \beqa
 H(x^0) &=& H(t_R^{})\left(\frac{a(t_R^{})}{a(t)}\right)^{\frac{d}{2}}~, 
 \\
 \rho(x^0) &=& \rho(t_R^{})\left(\frac{a(t_R^{})}{a(x^0)}\right)^{d}~, \  
 \beta(x^0) = T_M^{-1}\frac{a(x^0)}{a(t_R^{})}~. 
 \eeqa
 If the conserved charge density $q(x^0) :=\rho(x^0) \beta(x^0)$ is identified with the entropy density for radiations as 
 \beqa
 \rho(x^0) \beta(x^0) &=&  {\rho(x^0) + P(x^0) \over T_R(x^0)} = {d  \over d-1}{\rho(x^0)\over T_R(x^0)}~, 
 \eeqa
 we can read the relation between radiation temperature $T_R(x^0)$ and $\beta(x^0)$ as\footnote{
 Due to the direct transition from the matter era to the radiation era imposed by hand in our simplified model, 
 this definition of $T_R^{}(x^0)$ is different from the usual definition of radiation temperature $\rho(x^0)=\sigma_d^{}T^d(x^0)$ in cosmology. 
In particular, when the decay rate of inflaton $\Gamma_\phi^{}$ satisfies $\Gamma_\phi^{}\ll H_I^{}$, the ordinary reheating temperature is given by $T_R^{}:=T(t_R^{})\sim \sqrt{\Gamma_\phi^{}M_{\rm Pl}^{}}$, which is not related to $T_M^{}$.    
We expect that this mismatch will be fixed in more realistic models.  
 }
 \beqa
 \beta(x^0) = {d\over d-1} {1\over T_R(x^0)} \Rightarrow
 T_R(x^0) = {d\over d-1}  T_M^{}\frac{a(t_R^{})}{a(x^0)} 
 ~, ~~~~
 \eeqa
 which shows an extra factor $d/(d-1)$ between $\beta(x^0)$ and the inverse temperature. 
 Thus, while $\beta(x^0)$ is continuous, the matter temperature $T_M$ and the radiation temperature $T_R(t_R)$ at $x^0=t_R$ are different as $(d-1) T_R(t_R) = d T_M$, as in the case at $x^0=t_M$. 
   
 It is non-trivial and interesting that this time dependent temperature satisfies the Stefan-Boltzmann law as 
 \beqa
 \rho(x^0) &=& C_R^{}  T_R^{d}(x^0)  , \quad C_R:= 
 \rho(t_R^{})\left({d-1\over d T_M }\right)^d,~~
 \eeqa
  which again justifies our interpretation on $Q$  and $\beta(x^0)$. 
  Note that we have not assumed thermal equilibrium for radiation to obtain the above result.  

In addition, a starting time of radiation dominated era $t_R^{}$ can be estimated by assuming that $C_R$ reproduces the Stefan-Boltzmann constant for black-body radiations in $d$ dimensions~\cite{Cardoso:2005cd} as 
 \beqa
 C_R =\sigma_d :=  g_R^{} {(d-1) \Gamma(d/2) \zeta(d)\over \pi^{d\over 2}}~, 
 \eeqa
 where $g_R$ represents a number of degrees of freedom for radiations ($g_R=d-2$ for photons), and $\zeta(x)$ is the Riemann zeta function.
In such a case, we obtain
 \beqa
 t_R-t_M &=& {2\over d-1}\left[ \left({d-1\over d T_M}\right)^{d\over2}\sqrt{(d-1)(d-2) \over 16\pi G_N \sigma_d}-{1\over H_I}\right],~~~~~
 \eeqa
 and 
 a positivity of $t_R - t_M$ implies
 \beqa
 T_M^{d\over 2} < \sqrt{\pi} M_{\rm pl} \tilde T_H  \left({d-1\over d }\right)^{d\over 2} \sqrt{(d-1)(d-2) \over 4 \sigma_d},
 \eeqa
 which gives
 \beqa
 T_M^2 < {27\over 32}\sqrt{10\over \pi} M_{\rm pl} T_H \simeq 1.5  M_{\rm pl} T_H
 \eeqa
at $d=4$, where $M_{\rm pl} = 1/\sqrt{G_N}$ is the Planck mass.

\section{A comparison with a scalar model}
\label{eq:numerical}
Here we numerically solve the inflaton dynamics at  $d=4$ and compare it with the analytical results in the previous section. 
Equations are
\aln{
&\ddot{\phi}+(3H+\Gamma)\dot{\phi}+\frac{\partial V}{\partial \phi}=0~,
\\
&\dot{\rho}_R^{}+4H\rho_R^{}=\Gamma (\rho_\phi^{}+p_\phi^{})~,
\\
&
\dot{\beta}-3H\frac{P}{\rho}\beta=0~,\quad H^2=\frac{8\pi G_N}{3}\rho~,
\\
&\rho=\rho_\phi^{}+\rho_R^{}~,~\rho_\phi^{}=\frac{1}{2}\dot{\phi}^2+V(\phi)~,
\\
&P=p_\phi^{}+\frac{1}{3}\rho_R^{}~,~p_\phi^{}=\frac{1}{2}\dot{\phi}^2-V(\phi)~,
}
where $\Gamma$ represents the decay rate of inflaton. 
As for the inflaton potential, we choose 
\aln{V(\phi)=\frac{m_\phi^2}{2}\phi^2+\frac{\lambda}{4!}\phi^4
}
as a toy model though its CMB predictions are already excluded by Planck2018~\cite{Planck:2018jri}. 
We also introduce the radiation entropy 
\aln{
S=V_{3}a^{3}(x^0)\rho_R^{}(x^0)\beta(x^0)~,
}
which must coincide with $Q$ after the reheating. 

\begin{figure}
\begin{center}
\includegraphics[scale=0.5]{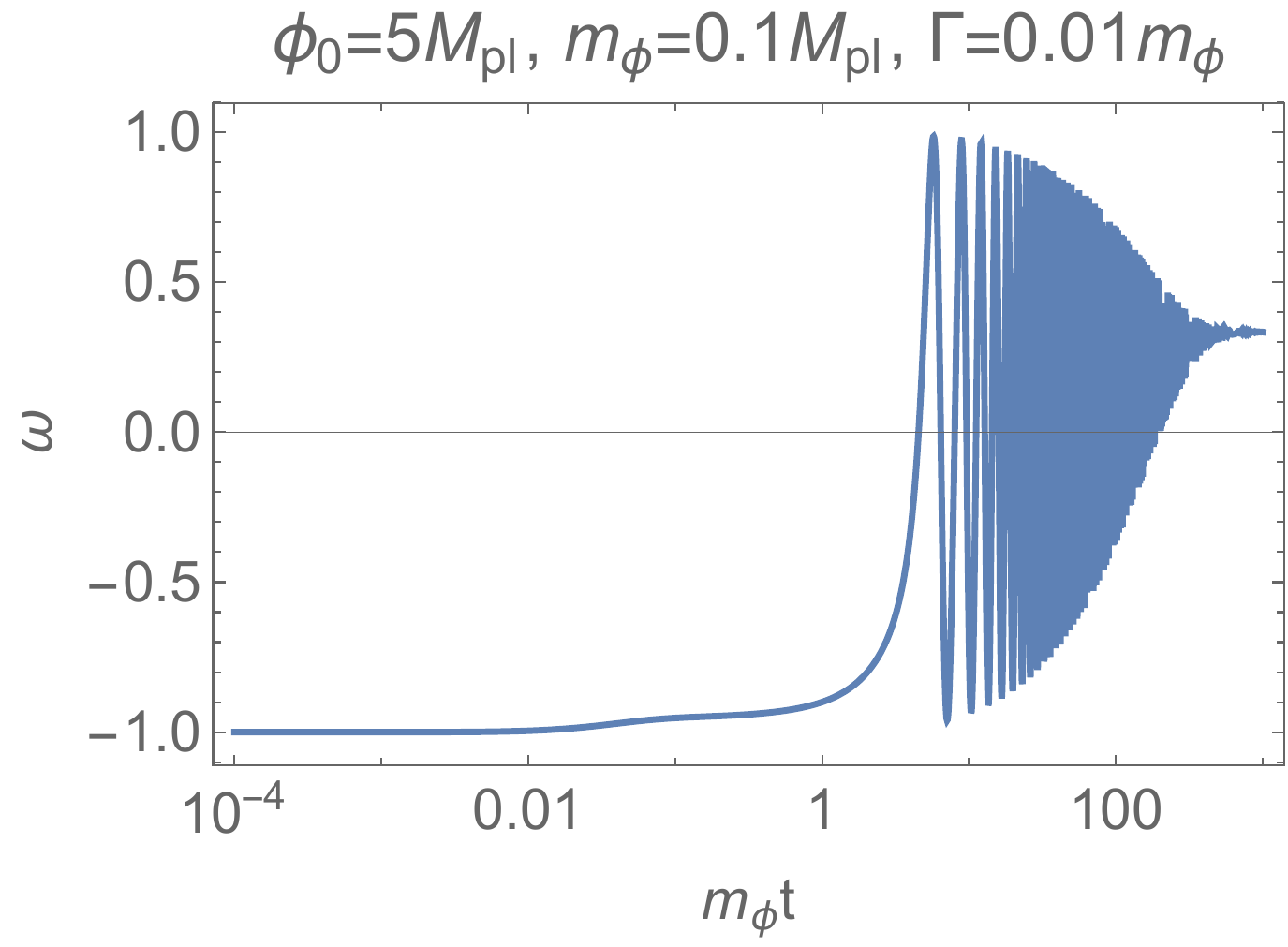}
\includegraphics[scale=0.5]{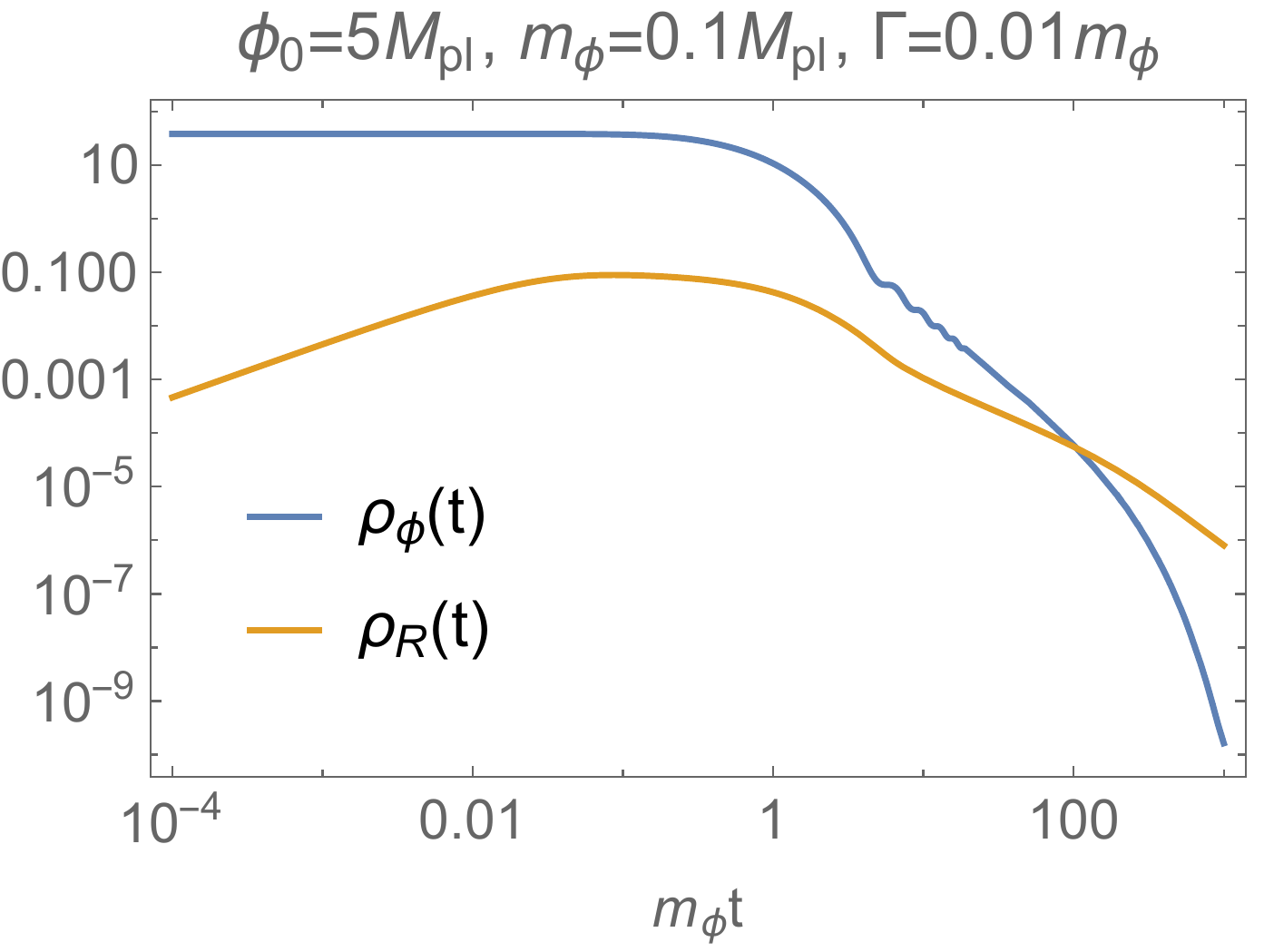}
\includegraphics[scale=0.5]{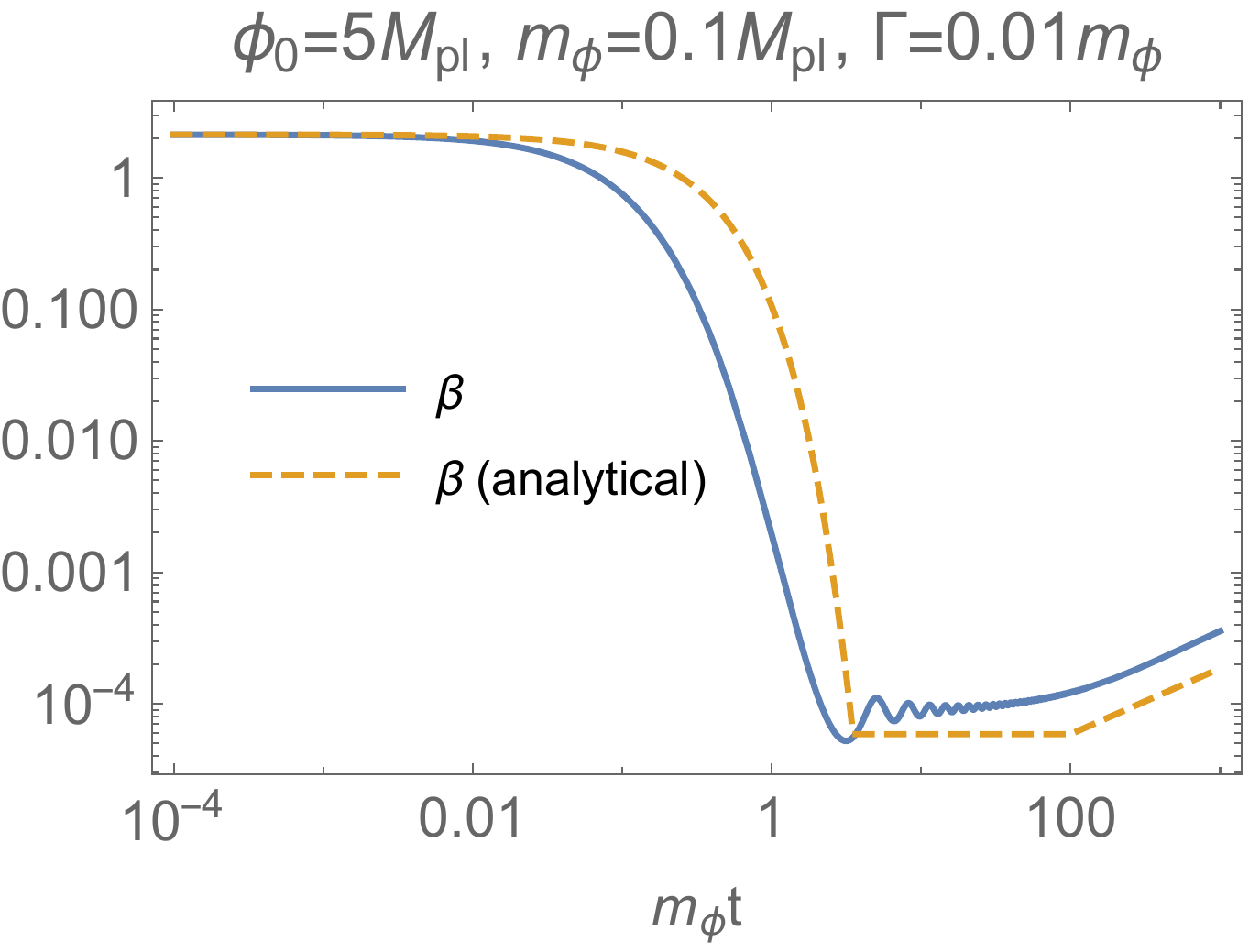}
\includegraphics[scale=0.5]{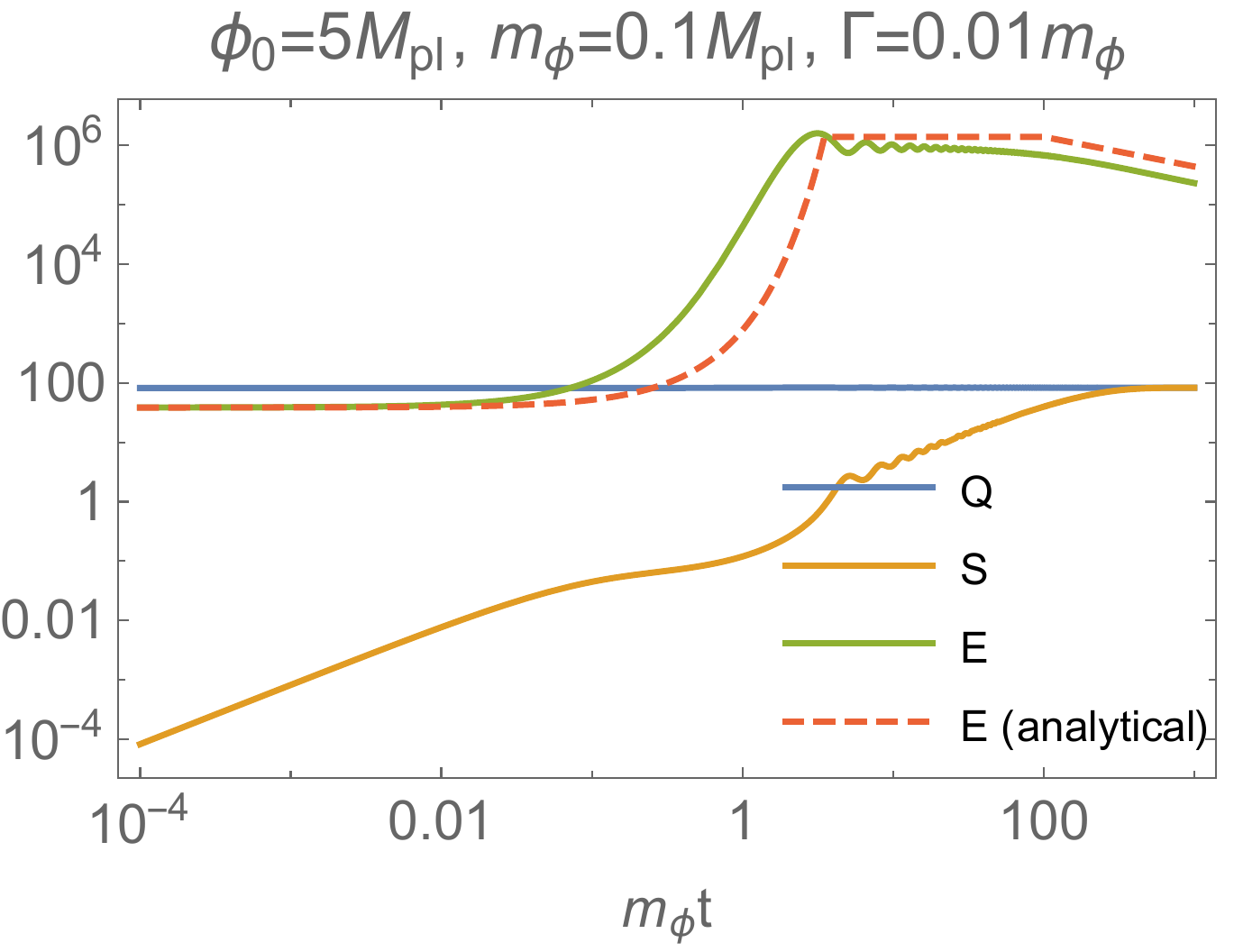}
\caption{(1st) EoS $\omega(x^0)$. (2nd) Energy densities $\rho_\phi$ (blue) and $\rho_R$(orange).
(3rd) The inverse temperature $\beta$ by analytic (orange) and numerical (blue) calculations. 
(4th) The conserved charge $Q$ (blue), the radiation entropy $S$ (orange) and energy $E$ (green) for this model, together with an analytic prediction of $E$ (red). 
}
\label{fig:result1}
\end{center}
\end{figure}

Numerical results are obtained with following parameters and initial conditions:
 \aln{
 \lambda &=10^{-2}~,\quad 
 m_\phi^{}=0.1M_{\rm Pl}^{}~,\quad \Gamma=10^{-2}m_\phi^{}~,
 \\
 \phi(0)&=5M_{\rm Pl}^{}~,\quad \dot{\phi}(0)=0~,\quad \beta(0)=T_H^{-1}~.
 }
 Fig.~\ref{fig:result1} (1st) shows the EoS $\omega(x^0)$ as a function of $m_\phi^{}t$  (logarithmic scale in the horizontal axis), which counts the number of oscillations of $\phi(x^0)$ around the origin. 
  The plot indicates that a slow rolling during inflation ends 
  and an oscillation in the matter  dominated era starts around $m_\phi t \simeq O(1)$, while
  it ends at $m_\phi t \simeq m_\phi^{}\Gamma^{-1} =100$. 
Fig.~\ref{fig:result1} (2nd) shows a log plot  of energy densities of matters $\rho_\phi$ and radiations $\rho_R$, which also shows that the radiation dominated era starts at  $m_\phi^{}t\sim 100$, where $\rho_R^{}$ becomes comparable to $\rho_\phi^{}$.

%

In Fig.~\ref{fig:result1} (3rd),  we compare the inverse temperature $\beta$ between the numerical calculation (blue) and the analytic result (orange).  
A qualitative behavior of the inverse temperature in the numerical calculation is well captured by the analytic prediction, whose constant behavior tells us that the matter dominated era starts at around $m_\phi t \simeq 3$ and ends at $ m_\phi^{} t \simeq 100$. 
The 4th panel of Fig.~\ref{fig:result1} shows $Q$ (blue), $S$ (orange), and total energy $E$ (green), together with the analytic prediction of $E$ (red)
for a comparison.
One  can see that $Q$ is conserved during the transition dynamics and that it is converted to the radiation entropy $S$ in the end. 
In addition, the analytic prediction of $E$ well describes the numerical result qualitatively. 
      %


\section{Summary}
\label{sec:summary}

We have investigated properties of a conserved charge obtained by a vector field $\xi^\mu=-\beta \delta^\mu_0$
during dynamical evolutions of a homogeneous and isotropic expanding Universe with several different constant equations of state.  
While the conserve charge in the inflation era with an appropriate choice of the initial temperature agrees with the Bekenstein-Hawking entropy for de Sitter spacetime, it is identified as the matter entropy with a constant temperature (energy) in the matter dominated era. 
Finally the charge becomes the radiation entropy and the time dependent temperature is shown to satisfy the Stefan-Boltzmann law during the radiation dominated era.  
As a concrete example of such a dynamical transitions, we have numerically studied a scalar (inflation) model with radiations and found that 
time evolutions of the conserved charge and temperature are qualitatively reproduced. 
Our results give strong evidences on the interpretation that the conserved charge in general relativity~\cite{Aoki:2020nzm} is indeed entropy of the system together with the spacetime dependent temperature.

\acknowledgments
This work is supported in part by  the Grant-in-Aid of the Japanese Ministry of Education, Sciences and Technology, Sports and Culture (MEXT) for Scientific Research (Nos.~JP22H00129) and by ``Program for Promoting Researches on the Supercomputer Fugaku''
 (Simulation for basic science: from fundamental laws of particles to creation of nuclei).
K.K. would like to thank Yukawa Institute for Theoretical Physics, Kyoto University for the support and the hospitality during his stay by the long term visiting program. 



\bibliographystyle{apsrev}
\bibliography{Bibliography}

\end{document}